\documentclass[10pt, twocolumn, aps, pre,showpacs]{revtex4}
\usepackage{natbib}
\usepackage{amsmath}
\usepackage{amsfonts}
\usepackage{amssymb}
\usepackage{bbold}
\usepackage{ulem}
\usepackage{cancel}
\usepackage{graphicx}
\usepackage{subfigure}
\usepackage{braket}
\usepackage{placeins}
\usepackage{color}
\usepackage{makecell}
\begin{document}
\title{Vortex nucleation in rotating Bose-Einstein condensates with density-dependent gauge potential}
\author{Ishfaq Ahmad Bhat$^1$, Thudiyangal Mithun$^2$, Bishwajyoti Dey$^1$}
\address{$^1$ Department of Physics, Savitribai Phule Pune University, Pune, Maharashtra, India, 411007\\
$^2$ Department of Mathematics and Statistics, University of Massachusetts, Amherst Massachusetts 01003-4515, USA}
\begin{abstract}
We study numerically the vortex dynamics and vortex-lattice 
formation in a rotating density-dependent Bose-Einstein 
condensate (BEC), characterized by the presence of 
nonlinear rotation. By varying the strength of nonlinear 
rotation in density-dependent BECs,
we calculate the critical frequency, $\Omega_{\text{cr}}$, 
for vortex nucleation both in adiabatic and  sudden external trap rotations.
The nonlinear rotation modifies the extent of deformation experienced 
by the BEC due to the trap and shifts the $\Omega_{\text{cr}}$ 
values for vortex nucleation. The critical frequencies 
and thereby, the transition to vortex-lattices in an adiabatic
rotation ramp, depend on conventional \textit{s}-wave 
scattering lengths through the strength of nonlinear rotation, 
$\mathit{C}$, such that $\Omega_{\text{cr}}(\mathit{C}>0) < 
\Omega_{\text{cr}}(\mathit{C}=0) < \Omega_{\text{cr}}(\mathit{C}<0)$. 
In an analogous manner, the critical ellipticity 
($\epsilon_{\text{cr}}$) for vortex nucleation during an 
adiabatic introduction of trap ellipticity ($\epsilon$) 
depends on the nature of nonlinear rotation besides
trap rotation frequency. The nonlinear rotation additionally 
affects the vortex-vortex interactions and the motion 
of the vortices through the condensate by altering the 
strength of Magnus force on them. The combined result 
of these nonlinear effects is the formation of the 
non-Abrikosov vortex-lattices and ring-vortex 
arrangements in the density-dependent BECs.
\end{abstract}
\maketitle
\section{Introduction:}
Soon after the successful realization of Bose-Einstein condensates (BECs) in 
alkali metal atoms, the study of quantized vortices 
and their dynamics in them gained an active interest 
\cite{Matthews_prl_1999, Madison_prl_2000, Madison_prl_2001, Hodby_prl_2001, Abosheer_sci_2001,  Recati_prl_2001, Sinha_prl_2001}. 
Quantum vortices in BECs and superfluid $^{4}\text{He}$ \cite{Donnelly_cup_1991} 
are the topological defects that arise in response to the 
external rotation \cite{Madison_prl_2000, Madison_prl_2001, Hodby_prl_2001, Abosheer_sci_2001},
 shaking a BEC \cite{Yukalov_jpcs_2016, Spencer_arxiv_2022} or even moving an obstacle 
through the condensate \cite{Sasaki_prl_2010, Kwon_pra_2015}.
The rotation in a BEC is induced  
by stirring the atomic cloud and can be either adiabatic 
or sudden (nonadiabatic). In either case the nucleation 
of vortices within the condensate occurs only when the 
trap rotation exceeds a certain minimum value called 
critical frequency, $\Omega_{\text{cr}}$. At this stage 
the vortices in BEC become energetically favorable.
For rotational frequencies greater than $\Omega_{\text{cr}}$, 
multiple vortices enter the condensate and eventually 
form a vortex-lattice. In nonadiabatic rotations, the critical 
rotation frequency, $\Omega_{\text{cr}}$, at which a 
single vortex exists stably at the minimum of the trap, 
decreases with the increasing strength of two- 
\cite{Fetter_rmp_2009, Dalfovo_rmp_1999, Ghosh_pra_2004} and three-body \cite{Mithun_pre_2013} 
interaction strengths. The change in $\Omega_{\text{cr}}$ is, 
however, marginal at large interaction strengths and 
$\Omega_{\text{cr}}\simeq 0.65 \omega_{\perp}$, where  
$\omega_{\perp}$ is the transverse trapping frequency, 
for the creation of single vortex in the ENS experiment 
\cite{Chevy_prl_2000}. On the other hand, in an adiabatic rotation 
ramp for a fixed trap ellipticity, conventional BECs with 
only short range \textit{s}-wave interactions form vortex 
lattices when the trap rotation frequencies are $\approx 
\omega_{\perp}/\sqrt{2}$ \cite{Madison_prl_2000, Madison_prl_2001, Hodby_prl_2001}. 
This transition from the vortex free state to the 
vortex-lattice structure occurs via dynamical 
instability \cite{Madison_prl_2001,Sinha_prl_2001} and is independent 
of the \textit{s}-wave interaction strength and 
longitudinal confinement. Likewise, vortex-lattices are 
also achieved by varying $\epsilon$ in a linear fashion 
at fixed trap rotation \cite{Madison_prl_2001, Parker_pra_2006}. 
Here the transition to the vortex-lattice structure 
is characterized by the critical ellipticity, 
$\epsilon_{\text{cr}}$, depending only on the 
trap rotation frequency.
Long range dipole-dipole interactions and trap 
geometry within the dipolar BECs are, however, 
found to alter these critical frequencies and 
ellipticities \cite{Cai_pra_2018, Dell_pra_2007, Bijnen_prl_2007, Bijnen_pra_2009, Prasad_prl_2019}. 
\par 
In addition to providing a flexible test-bed 
for numerous quantum phenomena, BECs are even 
noted for the experimental realization of artificial 
electromagnetism in them \cite{Dalibard_rmp_2011, Goldman_rpp_2014}. 
This has been made possible through the engineering 
of artificial gauge potentials in these ultracold condensates.
Induced electromagnetism in BECs has lead to the emergence 
of novel phenomena like density-dependent 
magnetism \cite{Edmonds_prl_2013}, exotic spin-orbit \cite{Lin_nat_2011} 
and spin angular momentum couplings \cite{Chen_prl_2018, Zhang_prl_2019}. 
However, the nature of gauge potentials in BECs  is 
typically static with no feedback between light and 
matter waves. In dynamical  gauge  fields, on the other hand,
 nonlinear feedback between matter and gauge fields 
 plays a role. This back-action is expected to enrich 
 the physics of atomic and nonlinear systems and  
 has been realized in experiments recently \cite{Clark_prl_2018, Gorg_nat_2019}.
\par 
Scalar and nonrotating BECs with density-dependent 
gauge potentials, known as \textit{Chiral} condensates, 
have been already studied in one-dimension for anyonic 
structures \cite{Keilmann_nat_2011}, chiral solitons \cite{Aglietti_prl_1996, Ishfaq_pre_2021,Dingwall_njp_2018,Dingwall_pra_2019} 
and collective excitations \cite{Edmonds_epl_2015} in them. 
More recently, strong density-dependent gauge potentials 
are even shown to realize chaotic collective dynamics in 
two-dimensional BECs \cite{Chen_njp_2022}.
The response of density-dependent BECs against the 
trap rotation in two-dimensions has been studied in Refs.  
\cite{Edmonds_pra_2020, Edmonds_pra_2021} where vortex ground states 
have been  simulated numerically. The density-dependent gauge
potential realizes an effective nonlinear rotation in BECs, 
thereby leading to non-Abrikosov vortex-lattices and ring vortex 
arrangements. In contrast to the Abrikosov vortex-lattices
in scalar BECs \cite{Campbell_prb_1979,Sato_pra_2007,Mithun_pra_2016}, the 
non-Abrikosov vortex-lattices in  BECs with density-dependent 
gauge potentials lack the hexagonal (or triangular) symmetry 
in the vortex arrangements \cite{Edmonds_pra_2020, Edmonds_pra_2021}.
\par 
The present paper studies the effects of nonlinear 
rotation on the dynamics of vortex lattice formation 
in these BECs within the scope of time-dependent 
numerical simulations. 
The studies of vortex nucleation in conventional 
\cite{Madison_prl_2000, Madison_prl_2001, Sinha_prl_2001, Tsubota_pra_2003, Kasamatsu_pra_2003} 
and dipolar \cite{Bijnen_prl_2007, Bijnen_pra_2009, Prasad_prl_2019} BECs reveal 
that the transition to the vortex-lattices in an adiabatic 
rotation ramp depends only on the rotation frequency and 
trap ellipticity but is independent of the \textit{s}-wave
interaction strength. Here we show that unlike in 
conventional and dipolar condensates, this transition 
depends on the short range \textit{s}-wave interactions in 
density-dependent condensates. The density-dependent BECs 
show similar responses against the adiabatic and sudden 
rotations such that the number and arrangement of the 
nucleated vortices varies with the strength of the nonlinear 
rotations. In addition to the energy considerations, 
the non-Abrikosov vortex-lattices and ring-vortex 
arrangements in these condensates are a result of the 
modifications in the Magnus force experienced by the 
vortices and the repulsive interactions between them. 
\par
The subsequent material is structured as follows. 
In Sec. \ref{sec:model} we introduce the model in 
the form of Gross-Pitaevskii equation including 
the density-dependent gauge  potential. 
This is followed by 
Sec. \ref{sec:numerics} where we present the possible 
routes to vortex nucleation through numerical 
investigations based on the Crank Nicholson method. 
The work is finally summarized in Sec. \ref{sec:concl}.
\section{Model}\label{sec:model} 
We consider a weakly interacting BEC of $N$ two-level atoms 
coupled by a coherent light-matter interaction due to an applied laser 
field. Within the rotating-wave approximation, such a BEC is described 
by the following mean-field Hamiltonian \cite{Dalibard_rmp_2011, 
Edmonds_pra_2020, Butera_jpb_2016}:
\begin{equation}\label{eq:ham}
    \hat{\mathcal{H}} = \left(\frac{\mathbf{\hat{p}}^2}{2m}+ V(\mathbf{r})\right)\otimes \mathbb{1} + \hat{\mathcal{H}}_{\text{int}} + \hat{\mathcal{U}}_{\text{MF}}
\end{equation}
where in the first term $\mathbf{\hat{p}}$ is the momentum operator, 
$V(\mathbf{r})= m(\omega^2_{x}x^2 + \omega^2_{y}y^2 + 
\omega^2_{z}z^2)/2$ is the trapping potential and $\mathbb{1}$ symbolizes 
the $2\times 2$ unity matrix. The mean-field interactions, 
$\hat{\mathcal{H}}_{\text{int}} = (1/2)\text{diag}[\Delta_{1},\Delta_{2}]$
with $\Delta_{i}=g_{ii}|\Psi_{i}|^2 + g_{ij}|\Psi_{j}|^2$ and
$g_{ij} = 4\pi \hbar^2 a_{ij}/m$  where $a_{ij}$ are the respective 
scattering lengths of the collisions between atoms in internal 
states $i$ and $j$ ($i, j = 1, 2$). Further, the light-matter 
interactions accounted by Refs. \cite{Goldman_rpp_2014,Butera_jpb_2016},
\begin{equation}\label{eq:lm}
    \hat{\mathcal{U}}_{\text{MF}} = \frac{\hbar \Omega_{r}}{2}
        \begin{pmatrix}
        \text{cos}\theta\left(\mathbf{r}\right)& e^{-i\phi\left(\mathbf{r}\right)}\text{sin} \theta\left(\mathbf{r}\right) \\
        e^{i\phi\left(\mathbf{r}\right)}\text{sin} \theta\left(\mathbf{r}\right) & -\text{cos} \theta\left(\mathbf{r}\right)
        \end{pmatrix}
\end{equation}
are parameterized in terms of Rabi frequency, $\Omega_{r}$,
mixing angle, $\theta\left(\mathbf{r}\right)$ and phase of the 
incident laser beam, $\phi\left(\mathbf{r}\right)$. 
We use perturbation theory by defining the perturbed states in terms of 
the unperturbed eigen-states $\ket{\pm}$ of $\hat{U}_{\text{MF}}$
\cite{Butera_jpb_2016,Edmonds_pra_2020,Edmonds_pra_2021}:
\begin{equation}
    \ket{\Psi_{\pm}} = \ket{\pm} \pm \frac{\Delta_{d}}{\hbar \Omega_{r}}\ket{\mp}
\end{equation}
where $\Delta_{d} = \text{sin}\left(\theta/2\right) 
\text{cos}\left(\theta/2\right) \left(\Delta_{1}-\Delta_{2}\right)/2$
is the mean-field detuning. Since the qualitative details do not depend on
which of the two perturbed states is chosen, therefore, we chose the 
$\Psi_{+}$-state and write the effective Hamiltonian as \cite{Butera_jpb_2016,Edmonds_pra_2020,Edmonds_pra_2021}:
\begin{equation}\label{eq:ham+}
    \mathit{\hat{H}}_{+} = \frac{\left(\mathbf{\hat{p}}-\mathbf{A}_{+}\right)^2}{2m} + W_{+} + \frac{\hbar \Omega_{r}}{2} + \Delta_{+} + V (\mathbf{r}),
\end{equation}
where the vector gauge potential is defined as, 
$\mathbf{A}_{+}=i\hbar\braket{\Psi_{+}|\Psi_{+}}$, while the 
scalar potential, $W_{+}=\hbar^{2} |\braket{\Psi_{+}|\nabla \Psi_{-}}|^{2}/2m$ 
and $\Delta_{+}=(\Delta_{1}~\text{cos}^{2}(\theta/2) 
+ \Delta_{2}~\text{sin}^{2}(\theta/2))/2$ is the dressed mean-field interaction. 
The generalized mean-field Gross-Pitaevskii (GP) equation for $\Psi_{+}$ is finally obtained by extremizing the energy functional, 
$\mathcal{E}=\braket{\Psi_{+}|\hat{H}_{+}|\Psi_{+}}$ as
\cite{Edmonds_pra_2020,Butera_jpb_2016, Butera_njp_2016, Butera_pt_2017}:
\begin{equation}\label{eq:ggpe}
\begin{split}
    i\hbar \frac{\partial\Psi_{+}}{\partial t} &= \left[ \frac{\left(\mathbf{\hat{p}}-\mathbf{A}_{+}\right)^2}{2m} + \mathbf{a}_{1}\cdot\mathbf{J}\right]\Psi_{+} +\\
   & \left[V(\mathbf{r})+\frac{\hbar\Omega_{r}}{2}+2\Delta_{+}+W_{+}\right]\Psi_{+} +\\
   & \left[\mathfrak{n}_{+} \left(\frac{\partial W_{+}}{\partial \Psi^{*}_{+}} - \boldsymbol\nabla \cdot \frac{\partial W_{+}}{\partial \boldsymbol\nabla\Psi^{*}_{+}} \right) - \frac{\partial W_{+}}{\partial \boldsymbol\nabla\Psi^{*}_{+}} \cdot \boldsymbol\nabla \mathfrak{n}_{+}\right]
\end{split}
\end{equation}
Here $\mathfrak{n}_{+} = |\Psi_{+}(\mathbf{r},t)|^{2}$ is the atomic density of the BEC in the
$\ket{\Psi_{+}}$ state, $\mathbf{a}_{1} = \boldsymbol\nabla\phi \Delta_{d}\text{sin}\theta/
\mathfrak{n}_{+}\Omega_{r}$ is the strength of the coupling to 
the gauge field and 
\begin{equation}\label{eq:nlc}
    \mathbf{J} = \frac{\hbar}{2mi}\left[\Psi_{+}\left(\boldsymbol\nabla + \frac{i}{\hbar}\mathbf{A}_{+}\right)\Psi^{*}_{+}-\Psi^{*}_{+}\left(\boldsymbol\nabla -\frac{i}{\hbar}\mathbf{A}_{+}\right)\Psi_{+}\right]
\end{equation} 
is the nonlinear current. We now expand the gauge potentials $\mathbf{A}_{+}$ 
and $W_{+}$ to first order in small parameters \textemdash $\theta=\Omega_{r}/\Delta$ 
representing the ratio of Rabi frequency to laser detuning, 
and $\epsilon = \mathfrak{n}_{+}(\mathbf{r})(g_{11}-g_{12})/4\hbar\Delta$ 
which takes into account the collisional and coherent interactions, to obtain the 
simplified relations for the potentials \cite{Edmonds_pra_2020,Edmonds_pra_2021}: 
\begin{equation}\label{eq:vpot}
    \mathbf{A}_{+} = -\frac{\hbar \theta^{2}}{4}\left(1-4\epsilon\right)\boldsymbol\nabla\phi
\end{equation}
\begin{equation}\label{eq:spot}
\begin{split}
    \mathit{W}_{+}=& \frac{\hbar^{2}}{2}\bigg(\frac{\left(\boldsymbol\nabla\theta\right)^2
    \left(1-4\epsilon\right)+\theta^2\left(1+4\epsilon\right)
    \left(\boldsymbol\nabla\phi\right)^2}{4m} - \\
    & \boldsymbol\nabla\theta^2 \cdot\boldsymbol\nabla\epsilon\bigg)
\end{split}
\end{equation} 
We further invoke the spatial dependence in Rabi coupling through $\Omega_{\text{r}} = 
\kappa_{0} r$, where $r$ is the radial distance and $\kappa_{0}$ is a constant, and 
define the phase, $\phi= l \varphi$ where $l$ and $\varphi$ are respectively the 
angular momenta and polar angle of the incident laser beam \cite{Juzeliunas_pra_2005}. 
The density-dependent gauge potentials (Eqs. \eqref{eq:vpot} and \eqref{eq:spot}) 
when substituted in Eq. \eqref{eq:ggpe} in a rotating frame gives the
simplified three-dimensional GP equation as \cite{Edmonds_pra_2020,Edmonds_pra_2021}: 
\begin{equation}\label{eq:gpe3d}
    i \hbar \frac{\partial \Psi}{\partial t} = 
    \bigg[-\frac{\hbar^2}{2m}\nabla^2 +
    V(\mathbf{r}) -\Omega_{\mathfrak{n}}(\mathbf{r},t) \hat{L}_{z} +
    \text{g}_\text{eff} |\Psi|^2 \bigg] \Psi
\end{equation}
In Eq. \eqref{eq:gpe3d}, the subscript $+$ is omitted, 
the operator $\nabla^2 =\frac{\partial^2}{\partial x^2} + 
\frac{\partial^2}{\partial y^2} + \frac{\partial^2}{\partial z^2}$, and 
$\hat{L}_{z}  = -i \hbar \bigg( x \frac{\partial}{\partial y} - 
y \frac{\partial}{\partial x} \bigg)$ is the \textit{z} component 
of the angular momentum operator. The density-dependent rotation 
experienced by the BEC, $\Omega_{\mathfrak{n}}(\mathbf{r},t)$ has the form,
\begin{equation}\label{eq:om_rho}
\Omega_{\mathfrak{n}}(\mathbf{r},t) = \Omega_{0} + \mathit{C} \mathfrak{n}(\mathbf{r}, t),
\end{equation} 
where $\Omega_{0}$ represents the trap rotation and 
$\mathfrak{n}(\mathbf{r},t)= |\Psi(\mathbf{r},t)|^2$ is the number density 
of the BEC. The nonlinear rotation strength, 
$\mathit{C} =l \theta^{2}_{0}{(g_{11}-g_{12})}/( m \Delta)$ 
where $\theta_{0}=\kappa_{0}/\Delta$ while  the mean-field interaction,  
$\text{g}_{\text{eff}} = g_{11}+ \hbar (2l^{2}-1)\mathit{C}/{4l}$.
It is worth mentioning that the allowed values of $\mathit{C}$ 
in a density-dependent BEC with $\text{g}_{\text{eff}}>0$ are bounded by 
$\pm\mathit{C}_{max}$, where $\mathit{C}_{max} = 
\sqrt{8\pi (\text{ln} 2) \text{g}_{\text{eff}}/3 N m}$ \cite{Edmonds_pra_2021}. 
Otherwise the nonlinear rotation overcomes the trap and the BEC 
ceases to exist. As a consequence of the density-dependent gauge 
potentials, the nonlinear rotation, 
$\mathit{C}\mathfrak{n}(\mathbf{r},t)$ gives rise to the 
density modulated angular velocity within the BEC. 
The different density regions in a BEC with 
density-dependent gauge potentials, therefore, experience different 
rotations depending on the nature of nonlinear rotation. 
Such effective nonlinear rotations in the BEC can be 
induced by employing laser beams with spatially varying 
intensities and carrying $l$ units of orbital angular 
momenta. From an experimental point of view, one can 
employ Laguerre-Gaussian laser beams with cylindrically 
varying intensity profiles and $l=1$, recently used in realizing 
spin-angular-momentum-coupled BECs \cite{Chen_prl_2018}. 
Furthermore, the density-dependent rotation in Eq. \eqref{eq:om_rho} 
is the combined (rigid-body and nonlinear) rotation, and the 
second gauged rotation still exists even when the externally driven 
rotation is absent ($\Omega_{0}=0$). The larger the number of atoms in a BEC, 
the easier it is to induce the nonlinear rotation and, therefore,
the nucleation of the vortices in the BEC.
\par 
By expressing $\Psi(\mathbf{r},t) = \psi (\boldsymbol{\rho},t)
\text{exp}(-z^{2}/2\sigma^{2}_{z})/\sqrt[4]{\pi \sigma^{2}_{z}}$ 
for a BEC of thickness $\sigma_{z}$ confined in a highly 
anisotropic trap with $\omega_{z}>> \omega_{\text{x,y}}$, 
and following the usual procedure of dimensional reduction 
\cite{Yunyi_mt_2005}, Eq. \eqref{eq:gpe3d} results in the following 
two-dimensional (2D) GP equation:
\begin{equation}\label{eq:model}
i \hbar \frac{\partial \psi}{\partial t} = 
\bigg[-\frac{\hbar^2}{2m}\nabla^2_{\boldsymbol{\rho}} +
V(\boldsymbol{\rho}) -\Omega_{n}(\boldsymbol{\rho},t) \hat{L}_{z} +
\text{g} |\psi|^2 \bigg] \psi
\end{equation}
The BEC, because of the high anisotropy and the absence 
of axial modes in case of sufficiently weak interactions 
\cite{Petrov_prl_2000}, assumes a pancake geometry such that 
$\boldsymbol{\rho} = (x,y)$ represents the position vector 
in \textit{xy} plane. In quasi-2D settings, the operator 
$\nabla^2_{\boldsymbol{\rho}} =\frac{\partial^2}{\partial x^2} + 
\frac{\partial^2}{\partial y^2}$, $V(\boldsymbol{\rho}) = 
\frac{1}{2}m (\omega^2_{x} x^2 + \omega^2_{y} y^2)$ is the 
2D harmonic trapping potential, and the nonlinear coefficient, 
$\text{g} = \text{g}_{\text{eff}}/\sqrt{2\pi}\sigma_{z}$.
The density-dependent trap rotation $\Omega_{n}(\boldsymbol{\rho},t)$ 
has the same form as that in Eq. \eqref{eq:om_rho} except that 
$\mathit{C}$ is now scaled by a factor of $1/\sqrt{2\pi}\sigma_{z}$ 
and $n(\boldsymbol{\rho},t)= |\psi|^2$ is the 2D density of the BEC.
In an elliptical trap with $\omega^2_{\perp} = (\omega^2_{x}+\omega^2_{y})/2$ 
as the transverse trapping frequency, we assume 
$\omega_{x} = \omega_{\perp}\sqrt{1-\epsilon}$ and 
$\omega_{y} =\omega_{\perp} \sqrt{1+\epsilon}$ such that 
$\epsilon = (\omega^2_{x}-\omega^2_{y})/(\omega^2_{x}+\omega^2_{y})$ 
determines the trap ellipticity. This ellipticity in the trapping 
potential breaks the rotational symmetry which is crucial for 
the nucleation of vortices in BECs. By means of spatiotemporal scaling
\cite{Kasamatsu_pra_2003, Victor_pra_1998}, $t = \omega^{-1}_{\perp} t^{\prime}, 
(x,y) = a_{\perp}(x^{\prime},y^{\prime}),
~\text{and}~ \psi=\frac{\sqrt{N}}{a_{\perp}}\psi^{\prime}$,
Eq. \eqref{eq:model} can be recast in the following normalized form:
\begin{equation}\label{eq:dmless_eq}
(i- \gamma) \frac{\partial\psi}{\partial t} = 
\bigg[-\frac{\nabla^2_{\boldsymbol{\rho}}}{2} +
V(\boldsymbol{\rho}) -\tilde{\Omega}_{n}(\boldsymbol{\rho},t) \hat{L}_{z} + 
\tilde{\text{g}} |\psi|^2 \bigg] \psi.
\end{equation} 
In Eq. \eqref{eq:dmless_eq}, the scaled parameters,
$ \tilde{\text{g}} = \text{g} N m/\hbar^2$ and 
$\tilde{\Omega}_{n}(\boldsymbol{\rho},t) = \Omega_{0}/\omega_{\perp} +
 \tilde{\mathit{C}} n(\boldsymbol{\rho},t)$, where 
$\tilde{\mathit{C}} = \mathit{C}N m/(\hbar\sqrt{2\pi}\sigma_{z})$.
In this dimensionless form the 2D wavefunction $\psi(\boldsymbol{\rho},t)$ is now 
normalized to unity, the elliptical trapping potential, 
$V(\boldsymbol{\rho}) = \frac{1}{2}[(1-\epsilon) x^2 + (1+\epsilon)y^2]$ 
and $\hat{L}_{z} = -i\big(x \frac{\partial}{\partial y} - 
y\frac{\partial}{\partial x} \big)$ is the angular-momentum operator.
The term $\gamma$ is introduced to account for any dissipation 
in the system \cite{Choi_pra_1998}. By setting $\gamma=0.03$ 
\cite{Tsubota_pra_2003,Kasamatsu_pra_2003, Mithun_pra_2016, Mithun_pra_2014} and 
$\omega_{\perp} = 2\pi \times 200$ Hz in the following, 
we study the effects of density-dependent 
gauge potentials on the dynamics of vortex nucleation 
in a harmonically confined BEC by directly solving the generalised
GP equation \eqref{eq:dmless_eq} numerically. The dissipation term 
fastens the convergence to the equilibrium 
state characterized by a vortex-lattice and  does not 
affect the dynamics of the BEC.
\section{Numerical Analysis}\label{sec:numerics}
The introduction of trap rotation, $\Omega_{0}$ and 
anisotropy, $\epsilon$ have been employed both 
experimentally \cite{Madison_prl_2000,Madison_prl_2001} and 
numerically \cite{Tsubota_pra_2003, Kasamatsu_pra_2003, Parker_pra_2006} to 
evolve the harmonically confined BECs for the nucleation 
of vortices. In either case, the BEC is first deformed 
to an elliptical cloud, undergoes shape oscillations and 
is subsequently nucleated with vortices. The deformation of the condensate 
due to the trap rotation is parameterized by the factor 
$\alpha_{0} = \Omega_{0}(\braket{x^2 - y^2}/
\braket{x^2 + y^2})$ with $\braket{\cdots}$ representing 
the expectation values and was first obtained theoretically 
by Recati \textit{et al.} \cite{Recati_prl_2001} by employing a 
hydrodynamic formulation \cite{Recati_prl_2001, Sinha_prl_2001}. 
In this formalism the value of $\alpha_{0}$ in units of 
$\omega_{\perp}$ satisfies the cubic equation, $\alpha^{3}_{0}+
\left( 1 +2\Omega_{0}/\omega_{\perp}\right)
\alpha_{0} - \epsilon \Omega_{0}/\omega_{\perp} =0$, whose
solutions for a given free parameter ($\Omega_{0}$ or $\epsilon$), 
form branches that mark the routes toward the 
instability in the respective parameter space. 
The free parameter ($\Omega_{0}$ or $\epsilon$)
is varied adiabatically until either a dynamical instability is reached 
or the $\alpha$ branch back-bends \cite{Madison_prl_2001, Recati_prl_2001,Bijnen_pra_2009}. 
For $\alpha_{0}>0$, the BEC system is always dynamically unstable; 
otherwise the instability is caused by the branch back-bending.
In the current scenario, the spatial dependence of the 
rotation complicates the hydrodynamic formulation. The phase profile 
of a rotating density-dependent BEC now also depends 
locally on the density distribution within the BEC. 
We, therefore, study the effects of density-dependent 
gauge potentials on the BEC deformation ($\alpha$) 
and vortex nucleation in a harmonically trapped 2D 
BEC by employing the split-step Crank Nicholson method
\cite{Muruganandam_cpc_2009}. The modifications 
in the effective mean-field coupling constant due to 
the gauge potential are neglected. 
\par 
An initial state for $\tilde{\text{g}}= 420$ and  
$\tilde{\mathit{C}} = 0, \pm 5$ is simulated via imaginary 
time propagation such that $\Delta x = 0.05$ and 
$\Delta t = 0.001$ are the space and time steps respectively.
The initially simulated ground state is then followed in real 
time by ramping up either $\Omega_{0}$ or $\epsilon$ linearly in 
time, thereby marking different routes for nucleation of vortices.
The experimental tuning of interaction parameters, $\tilde{\text{g}}$ and
$\tilde{\mathit{C}}$ through Feshbach resonances 
\cite{Inouye_nat_1998,Blatt_prl_2011} also paves way for controlling 
the dynamics of the respective condensate in the $\alpha$-$\Omega_{0}$ 
and $\alpha$-$\epsilon$ parameter spaces. In any case, 
the number of atoms in the BEC is conserved by fixing the 
chemical potential after every time step. Moreover, the
results remain unchanged for a ground state simulated with 
$\tilde{\mathit{C}} = 0$, followed by introducing $\Omega_{0}$ 
or $\epsilon$ ramp and the nonlinear rotation simultaneously in real time. 
\par
\begin{figure}[!htb]
\includegraphics[width=0.48\textwidth]{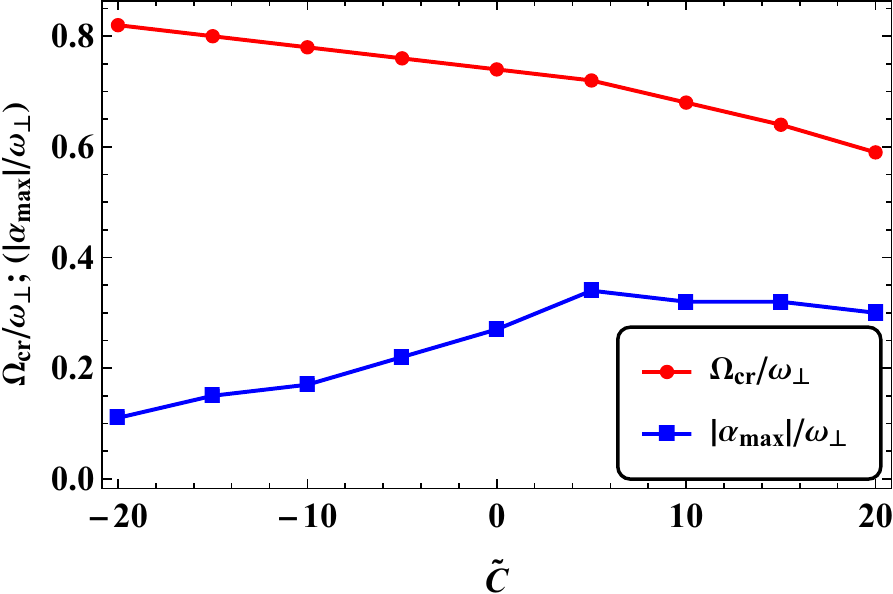}
\caption{Variation of the critical frequency, $\Omega_{\text{cr}}$, and  maximum deformation, $|\alpha_{max}|$ with the strength of nonlinear rotation, $\tilde{\mathit{C}}=\mathit{C} N m/(\hbar\sqrt{2\pi}\sigma_{z})$ in a harmonically confined BEC with $\tilde{g} = 420$.}\label{fig:om_alpha_var}
\end{figure}

 
We first consider the adiabatic increase of trap rotation
frequency by making it time-dependent, $\Omega_{0}(t) = ft$ for 
$f \approx 1.6\times 10^{-3}\omega^2_{\perp}$ and 
$\epsilon =0.025$. 
\begin{figure}[!htb]
\centering
\includegraphics[width=0.48\textwidth]{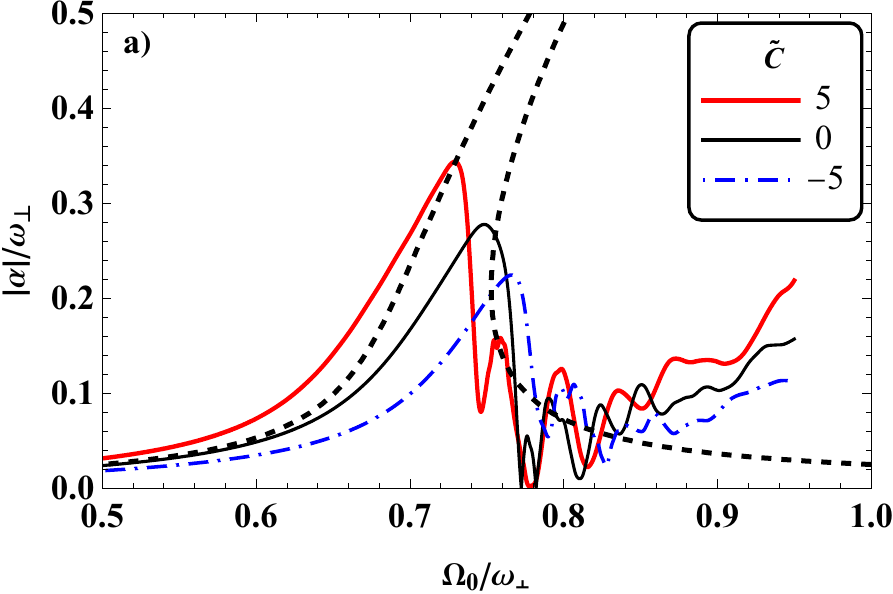}
\includegraphics[width=0.48\textwidth]{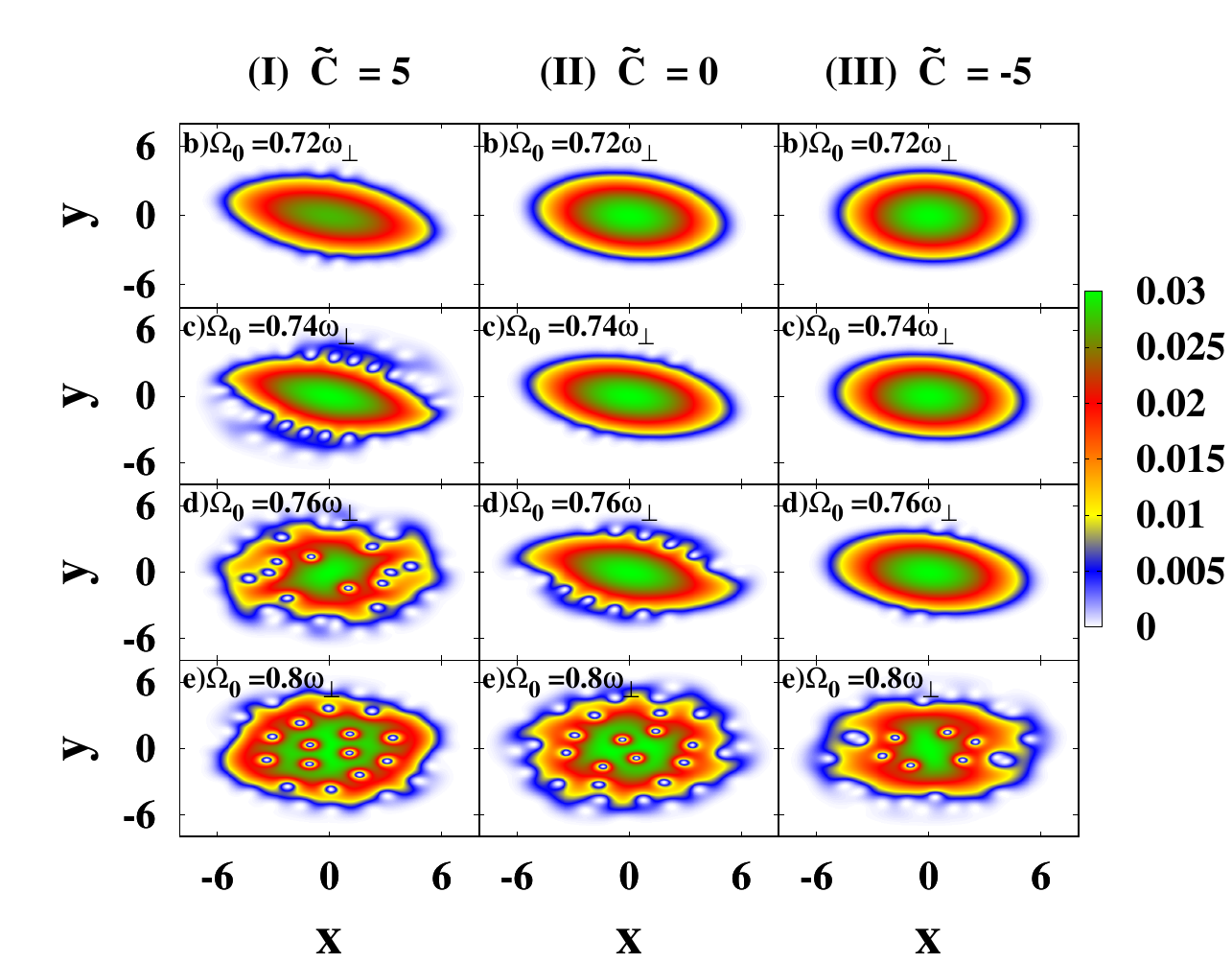}
\caption{(a) Time development of $|\alpha|/\omega_{\perp}$ for 
an ascending trap frequency, $\Omega_{0}(t)$ and $\epsilon=
(\omega^{2}_{x}-\omega^{2}_{y})/(\omega^{2}_{x}+\omega^{2}_{y}) = 0.025$. 
The color coded curves describe the numerical result for $\tilde{\mathit{C}}
=\mathit{C} N m/(\hbar\sqrt{2\pi}\sigma_{z}) = 5, 0 ~\text{and}~ -5$ 
respectively while the black dashed curves describe the analytical result for 
$\mathit{C} =0$ \cite{Recati_prl_2001, Sinha_prl_2001}. The discrepancy
between the analytical and numerical results for 
$\tilde{\mathit{C}}=0$ correspond to the deviation of 
the density profile from the Thomas-Fermi limit. 
Columns I, II and III represent the time evolution of 
condensate density with $\tilde{\mathit{C}} = 5, 0, ~\text{and}~ -5$ 
respectively.}\label{fig:om_ramp}
\end{figure}
The effect of the density-dependent gauge potentials on the 
deformation of the BEC is taken into account by defining,
\begin{equation}\label{eq:alpha}
\alpha=\frac{{\braket{\tilde{\Omega}_{n}(\boldsymbol{\rho},t)(x^2-y^2)}}}{{\braket{x^2+y^2}}}
\end{equation}
for $\tilde{\Omega}_{n}(\boldsymbol{\rho},t)=\Omega_{0}/\omega_{\perp}+ 
\tilde{\mathit{C}}|\psi(\boldsymbol{\rho},t)|^2$ and keeping the different 
parameter values same as mentioned in Ref. \cite{Madison_prl_2000}. 
The expressions $\braket{\cdots}$ in Eq. \eqref{eq:alpha} 
represent the expectation values and hence $\alpha$
represents the net average deformation experienced 
by the density-dependent BEC. As the rotation 
frequency, $\Omega_{0}$ is ramped up from zero, and
the condensate follows the left branch for $\alpha$ 
(See Fig \ref{fig:om_ramp}(a)), 
except for the small discrepancy due to its 
non-Thomas-Fermi nature, and becomes increasingly elongated 
as it goes through the higher values of $\alpha$. 
Once $\Omega_{\text{cr}}$ is reached, the elongation in 
the BEC is maximum and the nucleated vortices at the surface 
start entering the condensate 
because of dynamical instability \cite{Sinha_prl_2001, Parker_pra_2006}. 
The nucleated condensate finally returns to an axis-symmetric 
state specified by the drop in the $\alpha$ values. 
In Fig. \ref{fig:om_ramp} it is clear that the 
critical frequency values shift with the nature and 
strength of nonlinear rotation, $\tilde{\mathit{C}}$. 
For $\tilde{\mathit{C}} =\pm 5$, the critical 
frequencies are found to change by 
$\Delta\Omega_{\text{cr}} \approx \mp 0.02 
\omega_{\perp}$respectively against the value of 
$\Omega_{\text{cr}} \simeq 0.74 \omega_{\perp}$ 
for $\tilde{\mathit{C}} = 0$. In addition to shifting 
the critical frequency, the nonlinear rotation is seen 
to alter the values of $\alpha$ such that the 
condensates with larger $\tilde{\mathit{C}}$ values 
are more elongated while following the left branch. 
This is directly linked with the number of the 
nucleated vortices within the condensate and hence, 
angular momentum. Consequently, the BECs with 
$\tilde{\mathit{C}} > 0$ are densely populated 
with the vortices and have large angular momentum values
compared to the ones with $\tilde{\mathit{C}} < 0$ 
where the vortices mostly lie toward the periphery. 
This occurs due to the different rotations and hence 
energies ($E=E_{0}-\tilde{\Omega}_{n}(\boldsymbol{\rho})L_{z}$) associated 
with the varying condensate densities. The BECs with 
$\tilde{\mathit{C}} > 0$ have their lowest energy 
regions at the center and hence get populated with 
vortices profusely. On the other hand, density-dependent 
BECs with $\tilde{\mathit{C}} < 0$ have their lowest 
energy zones at the periphery, thereby populating the 
vortices around the condensate outskirts. The values 
of critical frequency, $\Omega_{\text{cr}}$ and the maximum 
condensate deformity, $|\alpha_{max}|$ in units of 
$\omega_{\perp}$ for the corresponding values of 
$\tilde{\mathit{C}}$ are shown in 
Fig. \ref{fig:om_alpha_var}.
It is clear that the shifts in critical frequency vary 
linearly with $\tilde{\mathit{C}}$ for 
$\tilde{\mathit{C}}<0$ and small $\tilde{\mathit{C}}>0$
values. For large $\tilde{\mathit{C}}>0$, 
the deviation from the linear behavior can be attributed to 
the non-Thomas-Fermi profiles of the corresponding BECs.
In an analogous manner, the maximum condensate deformations, 
$|\alpha_{max}|$ corresponding to the $\Omega_{\text{cr}}$ 
values increase with increasing $\tilde{\mathit{C}}$ and 
show anomaly for large values of $\tilde{\mathit{C}}>0$.
\par
\begin{figure}[!htb]
\includegraphics[width=0.48\textwidth]{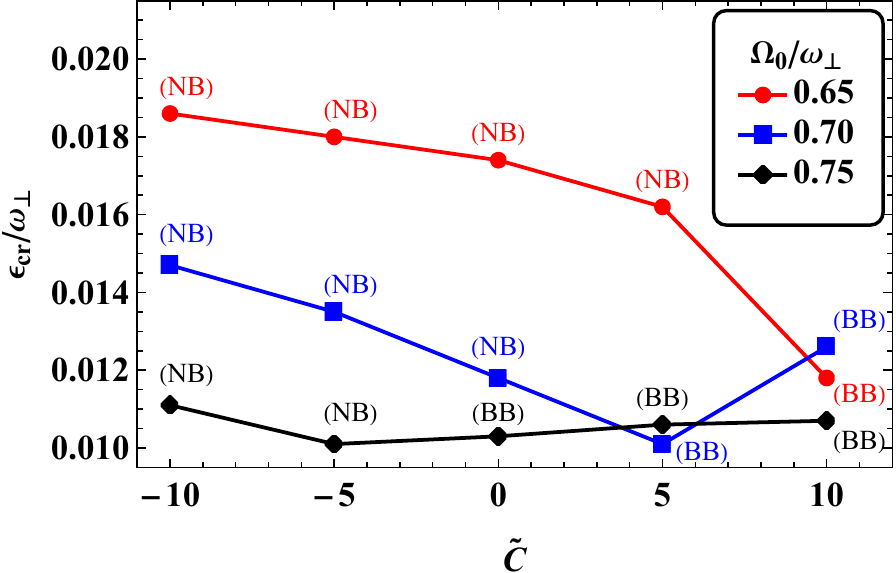}
\caption{Variation of $\epsilon_{\text{cr}}$ for corresponding values of $\Omega_{0}$ and $\tilde{\mathit{C}}=\mathit{C} N m/(\hbar\sqrt{2\pi}\sigma_{z})$ in a BEC with $\tilde{g}= 420$ and following an $\alpha$ branch with (BB) or without (NB) back-bending. }\label{fig:epsilon_var}
\end{figure}

 
Next, we consider the adiabatic increase in trap 
anisotropy by allowing $\epsilon$ to ramp up linearly 
from 0 to 0.025 at a rate $d\epsilon/dt \approx 
6.2\times10^{-5} \omega_{\perp}$ for a fixed trap 
rotation frequency, $\Omega_{0}$. Like in the case of 
$\Omega_{0}$ ramp, the BEC follows the respective 
$\alpha$ branch until the vortices start nucleating within 
the BEC at $\epsilon=\epsilon_{\text{cr}}$. In the ENS experiment 
involving an $\epsilon$ ramp \cite{Madison_prl_2001}, 
the nucleation of vortices in BECs with $\tilde{\mathit{C}} = 0$ 
was observed only if $\Omega_{0}>\omega_{\perp}/\sqrt{2}$ 
whence the $\alpha$ branch shows back-bending. 
However, for $\Omega_{0}<\omega_{\perp}/\sqrt{2}$, 
$\alpha$ is a monotonic function of $\epsilon$ and 
the numerical simulations \cite{Parker_pra_2006} suggest that 
vortex nucleation can occur at a larger $\epsilon$ 
due to ripple instability. Since the net rotation and 
deformation experienced by the BEC is altered by the 
presence of density-dependent gauge potentials.
Therefore, for a given $\Omega_{0}$, the nature of nonlinear 
rotation determines the branch followed by the BEC in 
$\alpha$-$\epsilon$ space such that for large values of 
$\tilde{\mathit{C}}>0$, the $\alpha$ branches exhibit 
back-bending even if $\Omega_{0}<\omega_{\perp}/\sqrt{2}$. 
Similarly with $\Omega_{0}\ge \omega_{\perp}/\sqrt{2}$, the 
$\alpha$ branches show no back-bending for large magnitudes 
of $\tilde{\mathit{C}}<0$. Consequently, this regulates the 
values of $\epsilon_{\text{cr}}$ for vortex nucleation.
\begin{figure}[!htb]
\centering
\includegraphics[width=0.48\textwidth]{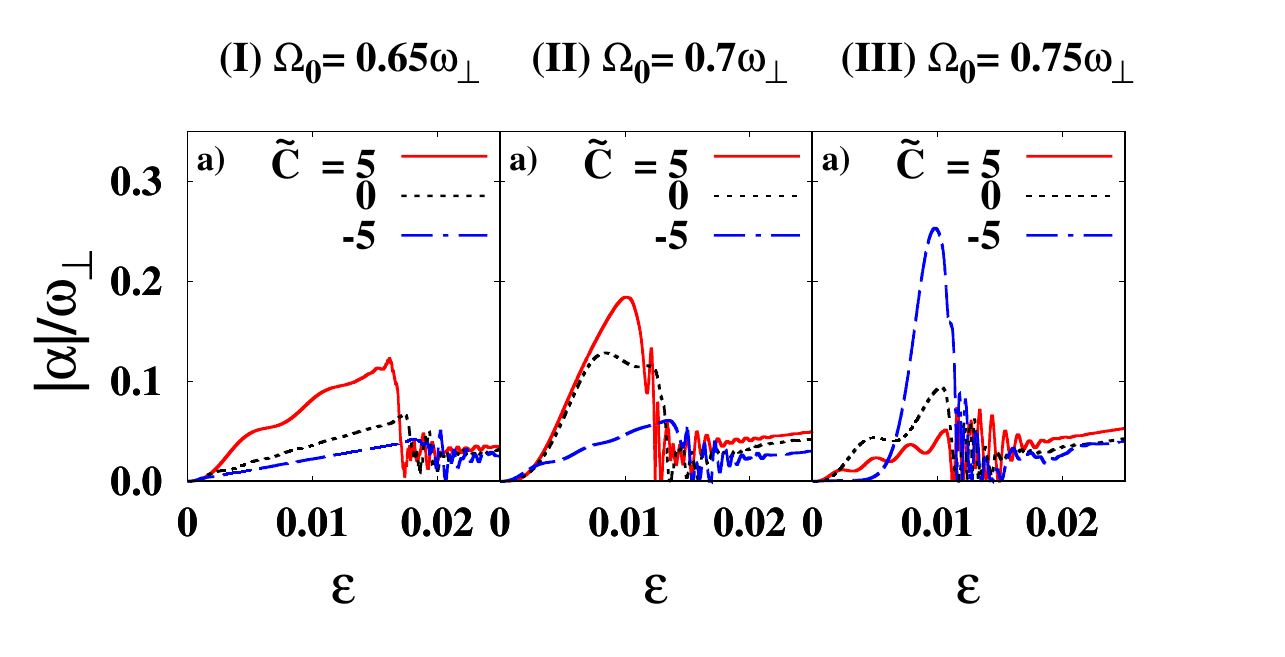}
\includegraphics[width=0.48\textwidth]{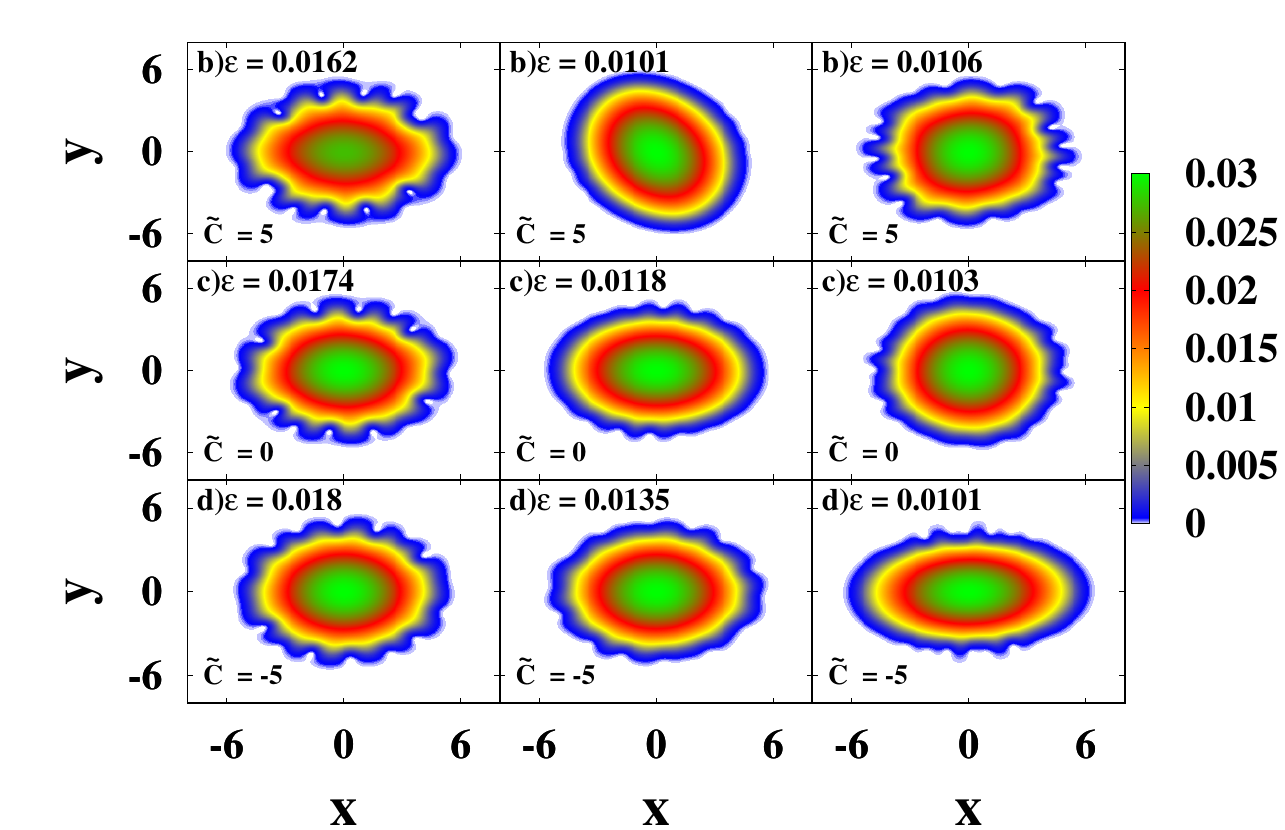}
\caption{(a) Time development of $|\alpha|/\omega_{\perp}$ 
for an ascending trap anisotropy, $\epsilon(t)$ at fixed 
$\Omega_{0}$ values. For $\Omega_{0}/\omega_{\perp} = 
0.65, 0.7, 0.75$, columns I (b)-(d), II (b)-(d) and III (b)-(d) 
respectively represent the density profiles  at mentioned 
$\epsilon_{\text{cr}}$ and $\tilde{\mathit{C}}$ 
values.}\label{fig:ep_ramp}
\end{figure}
Figures \ref{fig:ep_ramp} and \ref{fig:epsilon_var}
show that for a BEC with given value of 
$\tilde{\mathit{C}}$ and following an $\alpha$ branch 
without (with) back-bending, $\epsilon_{\text{cr}}$ 
decreases (increases) with the increase of the trap 
frequency. A similar effect due by the nonlinear 
rotation is inferred such that 
$\epsilon_{\text{cr}}(\mathit{C}>0) < 
\epsilon_{\text{cr}}(\mathit{C}=0) < 
\epsilon_{\text{cr}}(\mathit{C}<0)$ for a given value of 
$\Omega_{0}$ and the BEC follows an 
$\alpha$ branch without back-bending. 
The numerically obtained $\epsilon_{\text{cr}}$
values as a function of $\Omega_{0}$ and 
$\tilde{\mathit{C}}$ as well as the nature of 
the $\alpha$ branch followed by the BEC 
during an $\epsilon$ ramp are presented in 
Fig. \ref{fig:epsilon_var} whereby it is evident that 
for an $\alpha$ branch without back-bending, the
$\epsilon_{\text{cr}}$ values decrease with
$\tilde{\mathit{C}}$ values for fixed $\Omega_{0}$.
However, for a BEC following an $\alpha$ branch 
with back-bending, $\epsilon_{\text{cr}}$ increases
with $\tilde{\mathit{C}}$ at a given value of 
$\Omega_{0}$.
This is due to the drastic change in the BEC deformation by 
the presence of density-dependent gauge potentials and 
it depends  on which side of  $\omega_{\perp}/\sqrt{2}$, 
$\Omega_{0}$ lies. For $\Omega_{0} < \omega_{\perp}/\sqrt{2}$, 
the condensate deformation, $\alpha$, decreases together 
with $\tilde{\mathit{C}}$ while the situation is reversed 
with $\Omega_{0} > \omega_{\perp}/\sqrt{2}$, where the 
condensate deformation increases with decreasing 
$\tilde{\mathit{C}}$ values. This is nonetheless true for  
a narrow range of $\tilde{\mathit{C}}$ values and anomalies 
occur at larger magnitudes. In an $\epsilon$ ramp, 
the vortex nucleation in BECs with $\mathit{C}=0$ 
is shown to occur either by ripple, interbranch or 
catastrophic instability mechanisms, depending on the 
value of $\Omega_{0}$ with respect to 
$\omega_{\perp}/\sqrt{2}$ \cite{Parker_pra_2006}. 
In case of interbranch and catastrophic instability mechanisms, 
the respective $\alpha$ branches show back-bending in the 
$\alpha-\epsilon$ parameter space. However, the back-bending of 
the $\alpha$ branch in catastrophic instability occurs at a larger 
$\epsilon$ compared to that of the interbranch instability.
Though the nature of  the $\alpha$ branch followed by the BEC during 
an $\epsilon$ ramp is modified by nonlinear rotation. 
However, in our numerical simulations, we did not observe the 
interbranch and catastrophic instability mechanisms 
and hence the role of $\tilde{\mathit{C}}$ in tuning 
these mechanisms for vortex nucleation. As shown in 
Fig. \ref{fig:ep_ramp}, the vortex nucleation in all 
cases is due to the generation of ripples at the 
condensate surface rather than due to the shedding 
of low-density areas and contortion in interbranch 
and catastrophic instability mechanisms respectively. 
\par
\begin{figure}[!htb]
    \centering
    \includegraphics[width=0.48\textwidth]{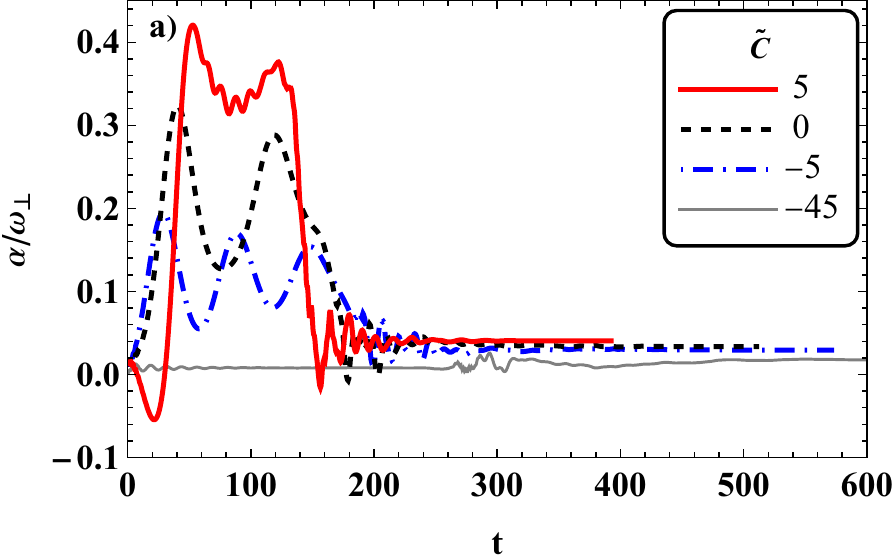}
    \includegraphics[width=0.48\textwidth]{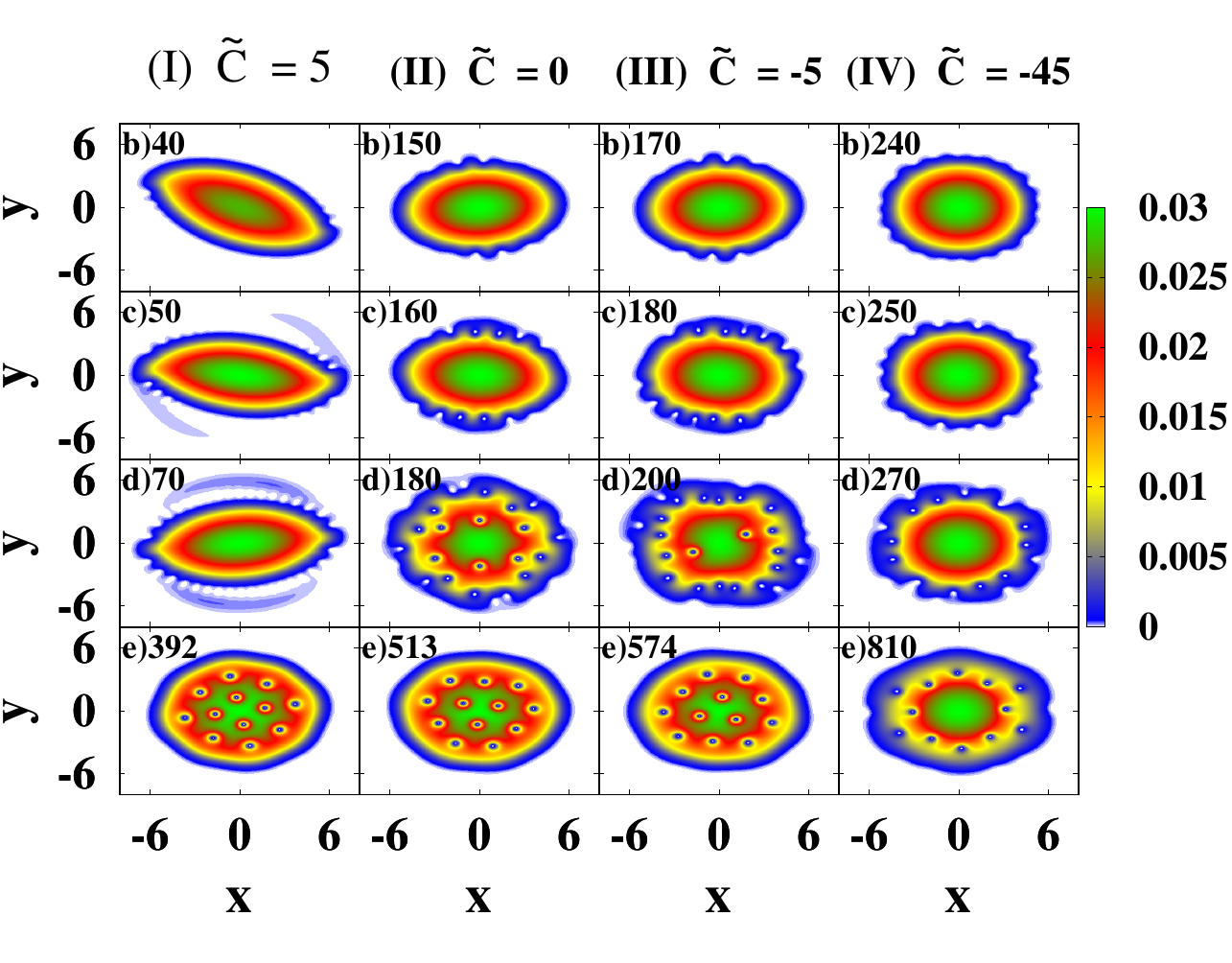}
    \caption{(a) Time development of $\alpha/\omega_{\perp}$ for different strengths of nonlinear rotation, $\tilde{\mathit{C}}$ in a BEC suddenly rotated with $\Omega_{0}=\omega_{\perp}/\sqrt{2}$. In the same scenario with $\tilde{g} = 420$, columns I-IV for $\tilde{\mathit{C}} = 5, 0, -5, -45$ respectively represent the time development of the BEC density.}
    \label{fig:sud_dy}
\end{figure}
\begin{figure}[!htb]
\includegraphics[width=0.48\textwidth]{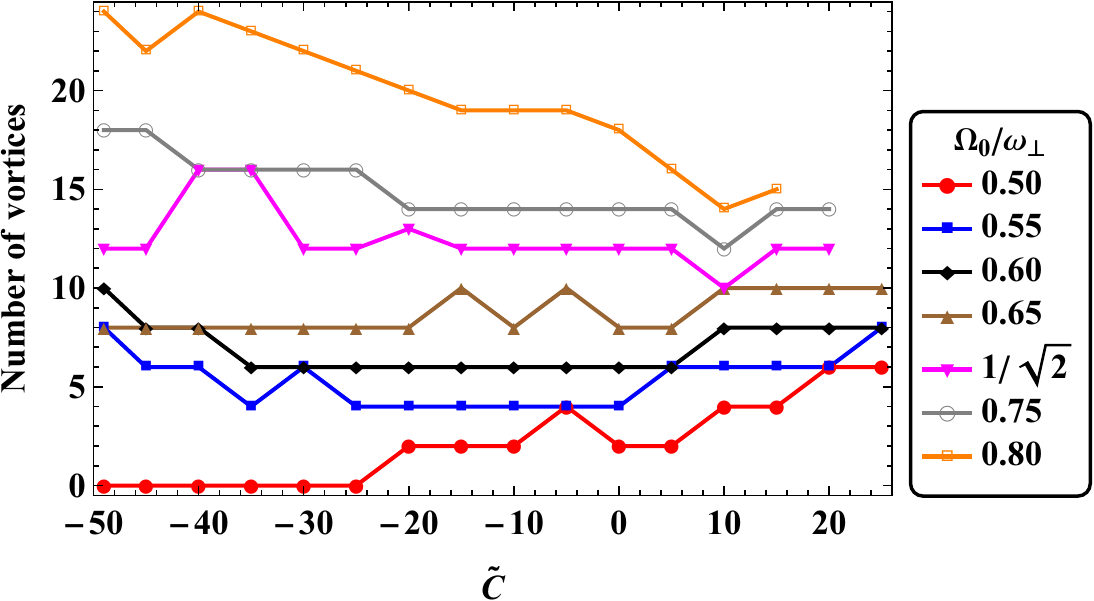}
\caption{Variation of the number of vortices with 
$\tilde{\mathit{C}}=\mathit{C}N m/(\hbar\sqrt{2\pi}\sigma_{z})$ for fixed values of 
$\Omega_{0}$ in a BEC with $\tilde{g} = 420$ and confined in an 
elliptical potential, $V(\mathbf{r}) = [(1-\epsilon) x^2 + 
(1+\epsilon)y^2]/2$ with $\epsilon=(\omega^{2}_{x}-\omega^{2}_{y})/
(\omega^{2}_{x}+\omega^{2}_{y})=0.025$}.\label{fig:vor_var_2f}
\end{figure}

So far we have focused on the response of the BEC with 
density-dependent gauge potentials to the slow turn on 
of the rotation of the trapping potential. We now consider 
the response of the BEC with the above set parameters 
against the sudden turn on of the rotation of the trap. 
When a BEC is rotated suddenly, it undergoes damped shape
oscillations and ultimately relaxes into a state with 
vortex-lattice \cite{Madison_prl_2000,Tsubota_pra_2003}. As shown in 
Fig. \ref{fig:sud_dy}, the nonlinear rotation has a 
prominent effect on these shape oscillations and thereby 
on the vortex nucleation, vortex dynamics and vortex-lattice 
formation. The shape oscillations become aperiodic and 
their amplitude increases for increasing 
$\tilde{\mathit{C}}>0$. On the other hand, the amplitude 
and time period of the shape oscillations decreases with
decreasing $\tilde{\mathit{C}}<0$ as shown in 
Fig. \ref{fig:sud_dy}(a). 
The nature of the vortex ground states and their 
achieving times are controlled by the nature and 
strength of nonlinear rotation, $\tilde{\mathit{C}}$. 
For $\tilde{\mathit{C}} =5$, the vortex ground state 
is achieved at short times and  the corresponding 
density profiles show that the BEC relaxes from the 
point of maximum elongation by shedding low-density 
material in a spiral pattern \textemdash 
\textit{fragmentation} \cite{Parker_prl_2005}. Consequently, 
surface waves and ghost vortices are generated which 
induce vortex nucleation. The vortices in the ground 
state, as shown in Fig. \ref{fig:sud_dy}(I)(e), 
crystallize into a non-Abrikosov lattice which lack the hexagonal
 symmetry in the vortex arrangements. The BEC becomes 
disordered for large values of $\tilde{\mathit{C}}>0$. 
In case $\tilde{\mathit{C}}=0$, perfect triangular 
lattices, as shown in Fig. \ref{fig:sud_dy}(II)(e), 
are obtained while the non-Abrikosov vortex
pattern again occurs for small values of $\tilde{\mathit{C}}<0$. 
The ring-vortex arrangements, shown in Fig. \ref{fig:sud_dy}(IV)(e), 
are obtained for large values of $\tilde{\mathit{C}}<0$. 
Moreover, the vortex nucleation in BECs with 
$\tilde{\mathit{C}} \leq 0$ occurs through ripples 
rather than shedding of low density regions in BECs 
with $\tilde{\mathit{C}} > 0$. In deformed BECs, 
the excitations occur on the surfaces with less 
curvature. Since the BEC deformation decreases 
together with  $\tilde{\mathit{C}}$, the condensates 
with large values of $\tilde{\mathit{C}}<0$ are much 
symmetric and excitations occur uniformly throughout 
the surface as shown in Figs. \ref{fig:sud_dy}(IV)(b) and 
\ref{fig:sud_dy}(IV)(c). With regard to the effect on the number 
of vortices in the equilibrium states due by the 
nonlinear rotation, Fig. \ref{fig:vor_var_2f} displays 
the variation of the number of vortices  with the 
strength of nonlinear rotation, $\tilde{\mathit{C}}$ 
for given values of $\Omega_{0}$. In the vicinity 
of $\Omega_{0}=\omega_{\perp}/\sqrt{2}$,
the number of vortices do not change for a larger domain 
of $\tilde{\mathit{C}}$ values, suggesting that the 
number of vortices in a suddenly rotated 
density-dependent BEC is fixed by the interaction strength, 
$\text{g}$, trap ellipticity, $\epsilon$, and rotation 
frequency, $\Omega_{0}$, while nonlinear rotation, 
$\tilde{\mathit{C}}n(\boldsymbol{\rho},t)$ only manipulates the 
position of the vortices within the condensate. However, 
this is not true for all trap rotation frequencies and 
the number of vortices also change on either side of
$\tilde{\mathit{C}} = 0$. Except at high trap rotations, 
the number of vortices change in multiples of two and 
is attributed to the twofold symmetry of the trapping potential.
\par
\begin{figure}[!htb]
    \centering
    \includegraphics[width = 0.48\textwidth]{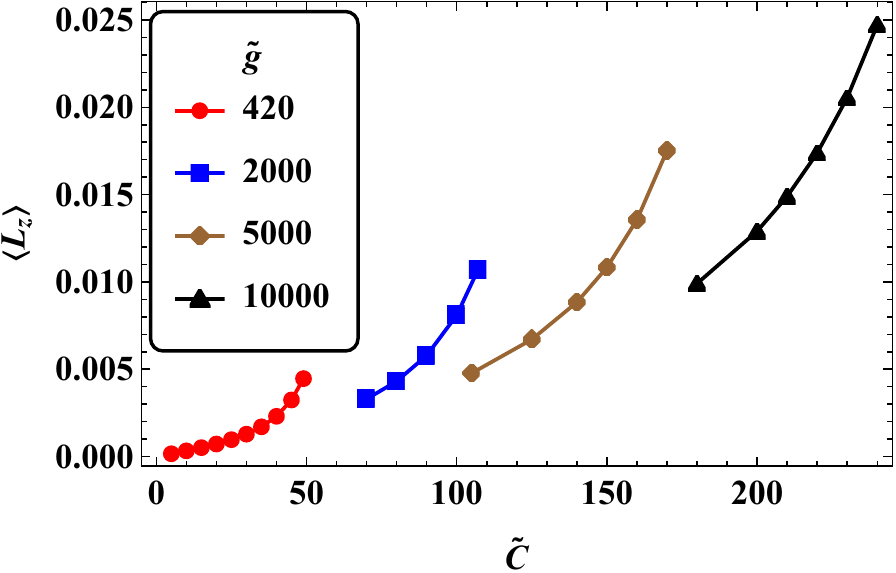}
    \caption{Variation of $\braket{L_{z}}$ with $\tilde{\mathit{C}}$ in the equilibrium states of the BEC for different strengths of $\tilde{g}$ and no external trap rotation, $\Omega_{0}=0$. The topmost data point corresponds to a value closer to $\tilde{\mathit{C}}_{\text{max}}$ for the respective case.}\label{fig:lz_vs_c}
\end{figure}

 
Further, it is worth mentioning that the vortex nucleation due to nonlinear 
rotation alone ($\Omega_{0} =0$) is anticipated for a BEC with large number of atoms \cite{Butera_jpb_2016, Butera_pt_2017}. Our numerical experiments along this line shown in
Fig. \ref{fig:lz_vs_c} displays that the average 
angular momentum, $\braket{L_{z}}= \braket{\psi|i(y\partial/\partial x - x\partial/\partial y)|\psi}$ values are still very small to ensure the 
formation of vortices within the BEC even for the mean-field interactions as large as, $\tilde{g} = 10\,000$. Neverthless, $\braket{L_{z}}$ increases with $\tilde{\mathit{C}}$ and $\tilde{g}$.
\par
We further examined the critical frequencies and the 
related single vortex dynamics by suddenly rotating 
the density-dependent BEC with  $\tilde{g} = 420$ in 
a harmonic trap displaced by $x_{0} = 0.3 a_{\perp}$ 
\cite{Kevrekidis_spr_2008}:
\begin{equation}\label{eq:pot_2}
V(\boldsymbol{\rho}) = \frac{1}{2}\bigg((x-x_{0})^2 + y^{2} \bigg)
\end{equation}
The critical frequencies in units of $\omega_{\perp}$ 
are found to be $0.443, 0.461, \text{and}~ 0.47$ for 
$\tilde{C} = 5, 0, \text{and} -5$ respectively. These 
results are in accordance with the already mentioned
results, though the BEC in this case deviates much from 
its stationary state because of sudden rotation.
\begin{figure}[!b]
    \centering
    \includegraphics[width=0.48\textwidth]{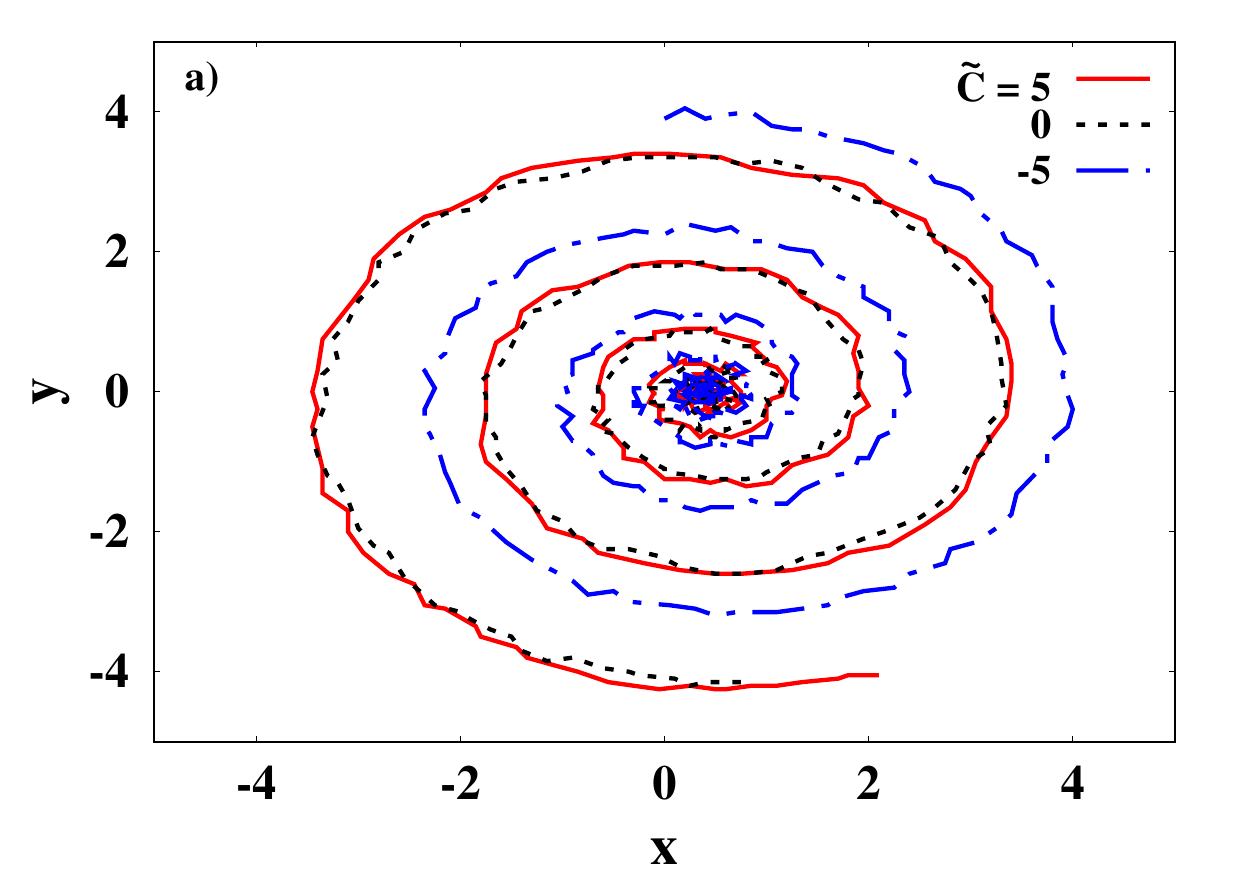}
    \includegraphics[width=0.48\textwidth]{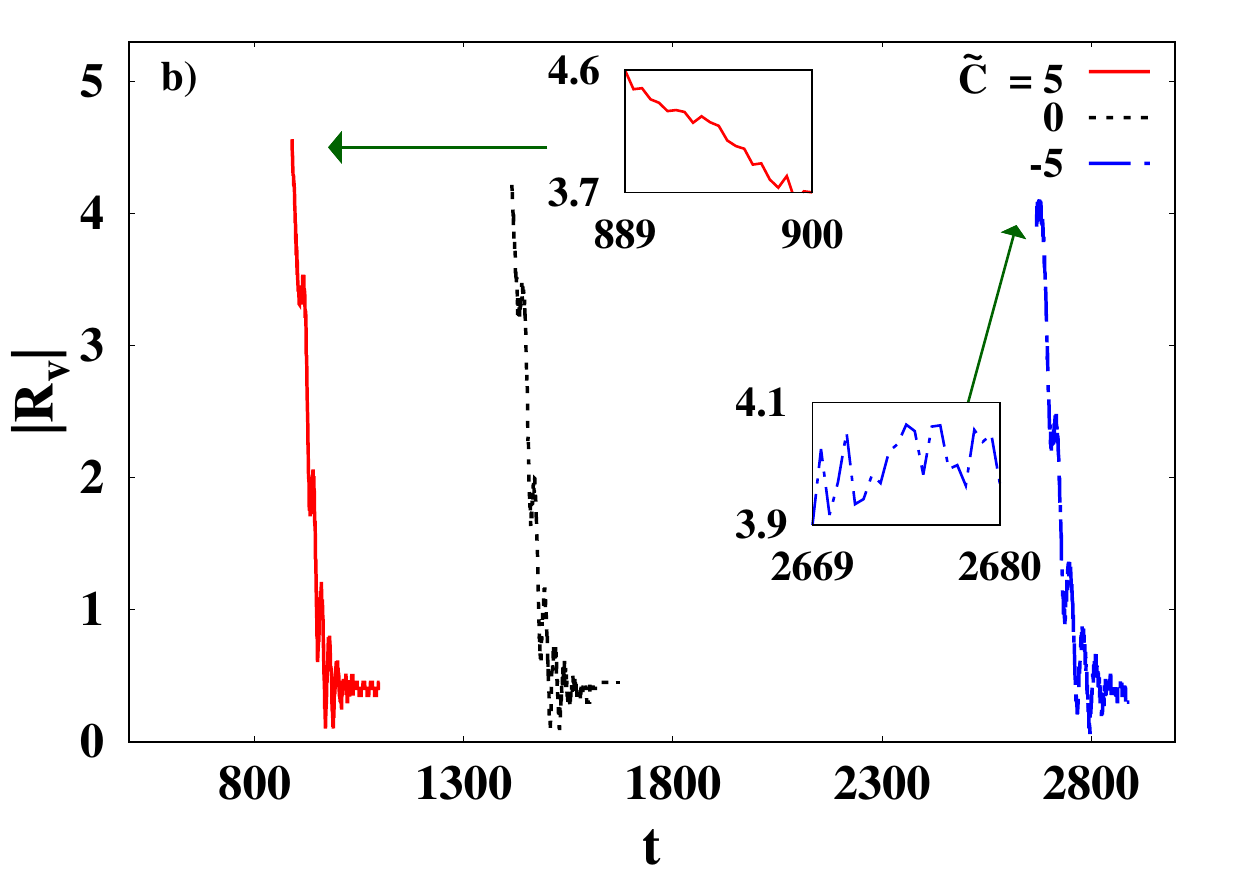}
\caption{(a) Single vortex trajectories in BECs  with 
$\tilde{g} = 420$, $\tilde{\mathit{C}} = 0, \pm5$ and 
rotated with $\Omega_{0} = 0.475 \omega_{\perp}$ in 
$xy$-coordinate space. (b) Time evolution of vortex 
position in terms of its radial distance $\mathbf{|R_{v}}|$ 
from the trap center. The vortex finally sits around 
the trap minimum at $(x_{0},y_{0}) = (0.3, 0.0)a_{\perp}$}
\label{fig:vor_track}
\end{figure}
The single vortex nucleated at the boundary of the BEC 
traverses a spiral path before settling down at the 
minimum of the trap. Because of the density-dependent 
gauge potentials, the single vortex for a fixed trap
rotation follows different trajectories 
depending on the nature of nonlinear rotation.
The effect of nonlinear rotation strengths on the 
trajectory of a single vortex is presented in 
Fig. \ref{fig:vor_track}, where vortex positions and 
its distance from the trap center, $|\mathbf{R_{v}}|$ 
are traced at different instants of time. For fixed 
$\Omega_{0} = 0.475 \omega_{\perp}$,  the single vortex 
first enters in BEC with  $\tilde{\mathit{C}}>0$, 
followed by the ones with 
$\tilde{\mathit{C}} = 0, \text{and} -5$ respectively. 
The first appearances of the vortex in the BECs with 
$\tilde{\mathit{C}} = 5, 0, -5$ were at 
$|\mathbf{R_{v}}|(a_{\perp}) \approx 4.6, 4.2, 3.9$.  
This is due to the larger net rotation and the associated size 
in case of BECs with $\tilde{\mathit{C}}>0$.  Moreover, 
in density-dependent BECs, the Magnus force, $\mathbf{f} = 
n(\boldsymbol{\rho},t)\mathbf{K} \times \mathbf{v}$ for a given density 
distribution, $n(\boldsymbol{\rho},t)$, circulation, $\mathbf{K}$ and 
vortex velocity, $\mathbf{v}= -y \tilde{\Omega}_{n}(\boldsymbol{\rho},t)
\hat{e}_{x} + x \tilde{\Omega}_{n}(\boldsymbol{\rho},t) \hat{e}_{y}$ is 
modified by the nonlinear rotation. In fact, the vorticity,
$\mathbf{\omega} = [2\tilde{\Omega}_{n}(\boldsymbol{\rho},t)+\tilde{\mathit{C}}
(x \partial n(\boldsymbol{\rho},t)/ \partial x + y \partial n(\boldsymbol{\rho},t)/ \partial y)] 
\hat{e}_{z}$ is now a function of both space and time.
The result is that in case of harmonically confined
density-dependent BECs with $\tilde{\mathit{C}}>0 $, 
the Magnus force is enhanced radially inwards and 
thereby the vortex moves swiftly toward the trap 
center, as confirmed in inset of Fig. 
\ref{fig:vor_track}(b). This is contrary to the 
case when $\tilde{\mathit{C}}<0$ where the Magnus 
force is reduced by the nonlinear rotation and 
hence the vortex moves slowly toward the trap center, 
as shown by the corresponding inset in 
Fig. \ref{fig:vor_track}(b). The vortex stays at the 
boundary for long before getting within the bulk of 
the condensate. The time taken in units of 
$\omega_{\perp}$ by the single vortex to reach 
the minimum of the trap is found to be 
$\approx 115, 186 ~\text{and}~ 221$  respectively 
for $\tilde{\mathit{C}} = 5, 0 ~\text{and} -5$. 
For large negative values of $\tilde{\mathit{C}}$, 
the Magnus force is outbalanced and the vortices 
reside at the fast rotating periphery, thereby favoring 
ring-vortex arrangements.  
\par
\begin{figure}[!htb] 
\centering
\includegraphics[width=0.49\textwidth]{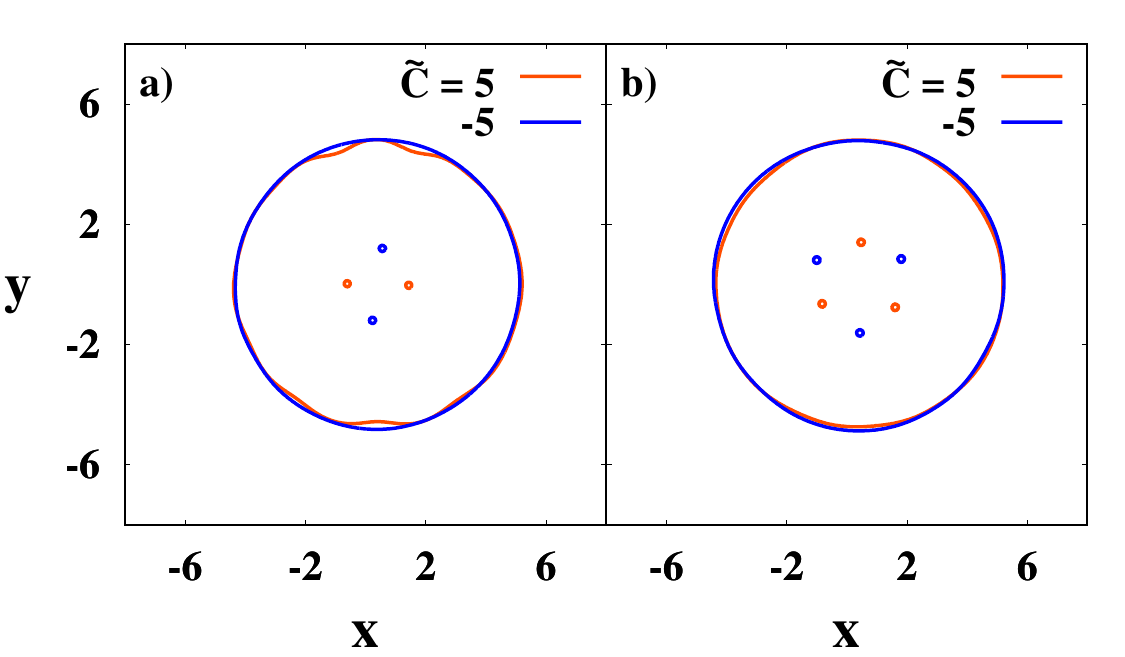}
\caption{Contour-plots showing vortices in density-
dependent BEC with $\tilde{\mathit{C}} = \pm5$ and 
rotated with (a) $\Omega_{0} = 0.51 \omega_{\perp}$ 
and (b) $\Omega_{0} = 0.525 \omega_{\perp}$ respectively. 
The respective results show same number of vortices with 
different positions for the same trap rotation 
frequency.} \label{fig:2_3_vortex}
\end{figure}

\begin{figure}[!htb]
\includegraphics[width=0.48\textwidth]{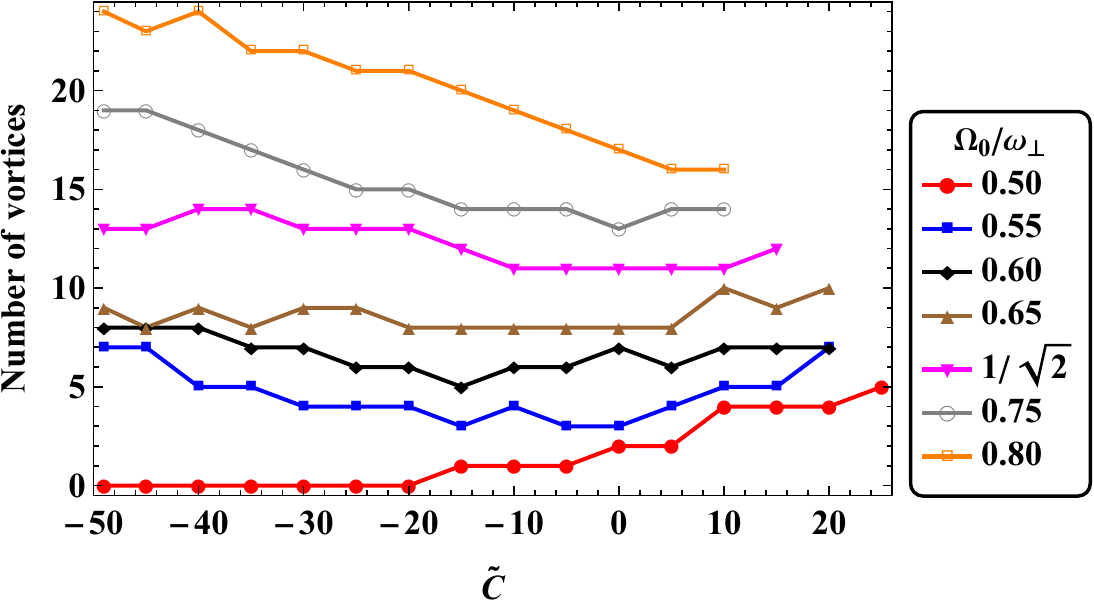}
\caption{Variation of the number of vortices with 
$\tilde{\mathit{C}} =\mathit{C}N m/(\hbar\sqrt{2\pi}\sigma_{z})$ for 
fixed values of $\Omega_{0}$ in a BEC with $\tilde{g} = 420$ and 
$V(\mathbf{r}) = ((x-x_{0})^2 + y^{2} )/2$.}\label{fig:vor_var_1f}
\end{figure}

 
The formation of ring-vortex arrangements is further 
supported by the presence of  modified repulsive 
interactions between the vortices within the BEC. 
In this direction we simulated the equilibrium states 
with two and three vortices respectively. It is found 
that the separation between the vortices increases 
with the strength of $\tilde{\mathit{C}}<0$ as 
shown in Fig. \ref{fig:2_3_vortex}. 
For $\tilde{\mathit{C}}>0$, the vortices come closer, making 
their separation smaller than $\tilde{\mathit{C}} =0$. 
Consequently, the strength of vortex-vortex 
repulsion decreases with the increase in the strength of 
nonlinear rotation and vice-versa. A condensate with the 
above set of values for the \textit{s}-wave interactions when
rotated at $\Omega_{0}=0.51\omega_{\perp}$ is nucleated 
with two vortices as shown in Fig. \ref{fig:2_3_vortex}(a).
The separation in units of $a_{\perp}$ between the two 
vortices for $\tilde{\mathit{C}} = 5 ~\text{and} -5$ 
are respectively found to be 2.05 and 2.41. 
These values correspond to an $\sim$ 8\% change 
in mean vortex separation with $\tilde{\mathit{C}} = 0$. 
The same results are inferred from the simulations 
of the ground states with three vortices shown in 
Fig. \ref{fig:2_3_vortex}(b), where the area changes 
by $\sim$ 15\% from the mean value. The change in separation 
between the vortices can be qualitatively explained by 
considering the method of images \cite{Verhelst_vdi_2017}. 
This method estimates that the equilibrium separation between 
two neighboring vortices in a BEC is $d = \frac{1}{\sqrt{2\Omega_{0}/
\omega_{\perp}}}$ as $\Omega_{0} \rightarrow \omega_{\perp}$. 
Along similar lines, for a density-dependent BEC 
$d \propto \frac{1}{\sqrt{\tilde{\Omega}_{n}(\boldsymbol{\rho},t)}}$ 
where $\tilde{\Omega}_{n}(\boldsymbol{\rho},t) = \Omega_{0}/\omega_{\perp} + 
\tilde{\mathit{C}} n(\boldsymbol{\rho},t)$.  
This relation shows that the distance between vortices decreases (increases) 
with the increasing (decreasing) strength of nonlinear 
rotation as shown in Fig. \ref{fig:2_3_vortex}. Moreover,
for a fixed value of $\tilde{\mathit{C}}$ the separation
between the vortices will also depend on the density 
distribution of the BEC. For a given value of $\Omega_{0}$ 
and $\tilde{\mathit{C}}>0$ the vortices in the center 
will be closer to each other than the ones in the periphery
of the BEC. However if $\tilde{\mathit{C}}<0$, then the 
vortices in the center will be distant from each other and 
for large negative values of $\tilde{\mathit{C}}$, the 
vortices reside only in the periphery of the BEC. The 
modifications in the vortex-vortex separation by the 
density-dependent gauge potentials thus result in the 
formation of non-Abrikosov vortex lattices and ring-vortex 
arrangements. Consequently, the areal density 
of vortices within the stationary states of 
density-dependent BECs is nonuniform and is regulated 
by the density distribution of the BEC through nonlinear 
rotation. These modifications due by the nonlinear rotation have an additional effect on the number of vortices within the BEC. Figure \ref{fig:vor_var_1f}
shows the variation of the number of vortices in the BEC 
trapped in a harmonic potential \eqref{eq:pot_2}  with 
$\tilde{\mathit{C}}$ for given values of $\Omega_{0}$. 
The results are consistent with those presented in 
Fig. \ref{fig:vor_var_2f}, except that the change in the 
vortex number now can be even or odd.
\section{Conclusions}\label{sec:concl}
In this work we consider a harmonically trapped 2D BEC 
subjected to a density-dependent gauge potential and 
theoretically studied the vortex nucleation in it by 
employing direct numerical simulations. The presence of
density-dependent gauge potential realises nonlinear 
rotation in BECs. The nonlinear rotation alters the net 
rotation and deformation experienced by the BEC. This 
consequently shifts the critical frequencies, $
\Omega_{\text{cr}}$ for the vortex nucleation. 
The shifts in the frequency values depend on the 
strength and nature of the nonlinear rotation. 
Consequently, the BECs with $\mathit{C}>0$ 
experience an increased net rotation and their 
critical frequencies are lowered. However, in BECs with 
$\mathit{C}<0$, the net rotation of the condensate is 
lowered while the critical frequencies soar higher. 
In general, for any BEC, the critical frequencies 
follow the order $\Omega_{\text{cr}}(\mathit{C}>0) < 
\Omega_{\text{cr}}(\mathit{C}=0) <
\Omega_{\text{cr}}(\mathit{C}<0)$. 
The critical frequencies for the transition 
to vortex-latices in BECs with density-dependent 
gauge potentials, therefore, depend on the strength of 
the short range \textit{s}-wave interactions within them,
contrary to the conventional and dipolar BECs with 
$\mathit{C} = 0$. The nonlinear rotation 
due to the density-dependent gauge potential also 
determines the number and arrangement of vortices within 
the rotating BECs. 
Moreover, in nonuniform BECs subjected to 
density-dependent gauge potentials, different 
density regions experience different rotations. 
This has a direct impact on the motion of the 
vortices within the condensate. In comparison to the
BECs with $\mathit{C} \leq 0$, it is found that 
a single vortex in BECs with $\mathit{C}>0$ 
moves swiftly and takes less time to reach the trap center 
when the system attains equilibrium. The Magnus force on the vortices  
is expected to get modified for the cause. Moreover, 
the mutual repulsive interactions between the 
vortices in a multiple vortex system are also 
modified by the presence of density-dependent 
gauge potentials. The distance between the vortices 
and their areal density now depends on the density 
distribution of the BEC and the strength and sign of 
nonlinear rotation. The combined effects of these 
modified forces and the interactions along with 
energy considerations result in non-Abrikosov 
vortex-lattices and ring-vortex arrangements.
\par 
Like critical frequency, nonlinear rotation due 
to the density-dependent gauge potentials also 
regulate the critical ellipticity, $\epsilon_{\text{cr}}$ 
for vortex nucleation. At a fixed $\Omega_{0}$, 
the critical ellipticities follow the trend of critical 
frequencies \textit{i.e}, 
$\epsilon_{\text{cr}}(\mathit{C}>0) < \epsilon_{\text{cr}}
(\mathit{C}=0) <\epsilon_{\text{cr}}(\mathit{C}<0)$ in case
the BEC follows an $\alpha$ branch without back-bending. 
However, the trend is reversed for a BEC following an 
$\alpha$ branch with back-bending.
\par 
It would be interesting to extend the present work
by considering density-dependent BECs in optical
lattices on top of the harmonic traps. 
In addition to the surface instability, vortex 
nucleation in conventional BECs with optical lattices
even occurs through the generation of vortex-antivortex 
pairs in the bulk \cite{Kato_pra_2011}. 
The pair generation occurs only if the lattice 
spacings are greater than the critical 
values. It is, therefore, natural 
to investigate the effects of density-dependent gauge 
potentials on the dynamics of vortex  nucleation in 
such BEC systems. Moreover, the findings can be 
realized in experiments since the density-dependent 
gauge potentials have already been demonstrated in 
two-dimensional optical lattices \cite{Clark_prl_2018, Gorg_nat_2019}.
\section{Acknowledgements}
The authors  thank P. G. Kevrekidis for fruitful discussions.
B.D. thanks Science and Engineering Research
Board, Government of India for funding through research 
project CRG/2020/003787. I.A.B. acknowledges CSIR, 
Government of India, for funding via CSIR 
Research Associateship (09/137(0627)/2020 EMR-I). 
\let\itshape\upshape\normalem
\bibliographystyle{apsrev}
\bibliography{references}

\end{document}